\documentclass[showpacs,twocolumn,prl]{revtex4-1}
\usepackage{epstopdf}
\usepackage{color}
\usepackage{bm}
\usepackage{graphicx}
\usepackage{amsmath}
\usepackage{amsfonts}
\usepackage{amssymb}
\usepackage{bezier} 
\usepackage{color}

\newcommand{\mib}{\bm}
 
\def\beq{\begin{equation}}
\def\eeq{\end{equation}}
\def\beq{\begin{equation}}
\def\eeq{\end{equation}}
\def\bea{\begin{eqnarray}} 
\def\eea{\end{eqnarray}} 

\def\frac#1#2{{\textstyle{#1 \over #2}}}

\def\half{\frac{1}{2}}

\mathchardef\sPi="7105
\mathchardef\sSigma="7106
\mathchardef\sPhi="7108
\mathchardef\sLambda="7103
\mathchardef\sOmega="710A
\mathchardef\sTheta="7102

\def\beq{\begin{equation}}
\def\eeq{\end{equation}}
\def \be{\begin{equation}}
\def \ee{\end{equation}}
\def \bea{\begin{eqnarray}}
\def \eea{\end{eqnarray}}
\def\half{\mbox{$1\over2$}}

\def \bS{{\bf S}}

\def \bd{{\bf d}}

\def \cC{{\cal C}}
\def \cS{{\cal S}}
\def\bOm{{\mib \Omega}}
\def\bfeta{{\mib \eta}}

\begin{document}
\title{$p6$ - Chiral Resonating Valence Bonds in the Kagom\'e Antiferomagnet}
\author{Sylvain Capponi$^{1}$}
\author{V. Ravi  Chandra$^{2,3}$}
\author{Assa Auerbach$^2$}
\author{Marvin Weinstein$^4$}
\affiliation{$^1$ Laboratoire de Physique Th\'eorique, Universit\'e de Toulouse and CNRS, UPS (IRSAMC), F-31062, Toulouse, France}
\affiliation{$^2$ Physics Department, Technion, Haifa 32000, Israel}
\affiliation{$^3$ School of Physical Sciences, National Institute of Science Education and Research,
Institute of Physics Campus, Bhubaneswar, 751005, India} 
\affiliation{$^4$ SLAC National Accelerator Laboratory, Stanford,  CA 94025, USA.}
\date{\today}
\begin{abstract}
The Kagom\'e Heisenberg antiferromagnet is mapped onto an effective Hamiltonian on  the star superlattice by Contractor Renormalization.
Comparison of ground state energies on large lattices to Density Matrix Renormalization Group justifies  truncation of effective interactions at range 3. 
Within our accuracy,  magnetic and translational symmetries are not broken (i.e. a spin liquid ground state).  However, we discover
doublet spectral degeneracies which signal the onset of  $p6$ - chirality symmetry breaking.This is understood by simple mean field analysis.
Experimentally, the $p6$ chiral order parameter should split the optical phonon  degeneracy near the zone center.
Addition of weak next to nearest neighbor coupling is discussed\footnote{This work was supported by the U. S. DOE, Contract No.~DE-AC02-76SF00515.}. \end{abstract} 
\pacs{75.10.Jm, 75.40.Mg}

\maketitle

The  antiferromagnetic Heisenberg model on the Kagom\'e lattice
\be {\cal H} = J\sum_{\langle ij\rangle} \bS_i\cdot\bS_j,~~~J>0,~S={1\over 2},
\label{H}
\ee 
is a much studied paradigm for frustrated quantum magnetism.  In
the classical approximation $S\to\infty$, this model exhibits
macroscopic ground state degeneracy which encumbers semiclassical
approximations. There is evidence, both numerical and experimental (in  ZnCu$_3$(OH)$_6$Cl$_2$~\cite{HS}), that quantum fluctuations
lead to a paramagnetic ``spin liquid" ground state~\cite{Balents2010}.

Exact diagonalization studies (ED)~\cite{ED},  and ED in the variational dimer singlets subspace~\cite{NNVB},
have not approached the thermodynamic limit, due to severe computer memory limitations.
Many methods have proposed 
paramagnetic ground states, including lattice symmetry breaking ``valence bonds crystals''~\cite{HVBC,MERA,QDM,var4}, algebraic spin liquids~\cite{var3} and a time reversal
symmetry breaking, chiral spin liquid~\cite{Messio}.

To date, the lowest energy on long cylinders has been found by Density Matrix Renormalization Group DMRG~\cite{DMRG1,DMRG2}.
The DMRG ground state is a translationally invariant singlet, with apparently no  broken translational or rotational symmetries. This state is consistent with a resonating
valence bonds (RVB) state~\cite{RVB} with a spin gap $\Delta_{S=1}=0.13$, (henceforth we express energies in units of $J$), and Z$_2$ topological
order~\cite{DMRG2,BalentsZ2}.  It is still unclear however, what are the
low-energy singlet excitations of this state~\cite{DMRG1,Jiang2008}, and whether or not any other
symmetry of ${\cal H}$ may be broken in the infinite two dimensional limit.
\begin{figure}
\includegraphics[width=\linewidth]{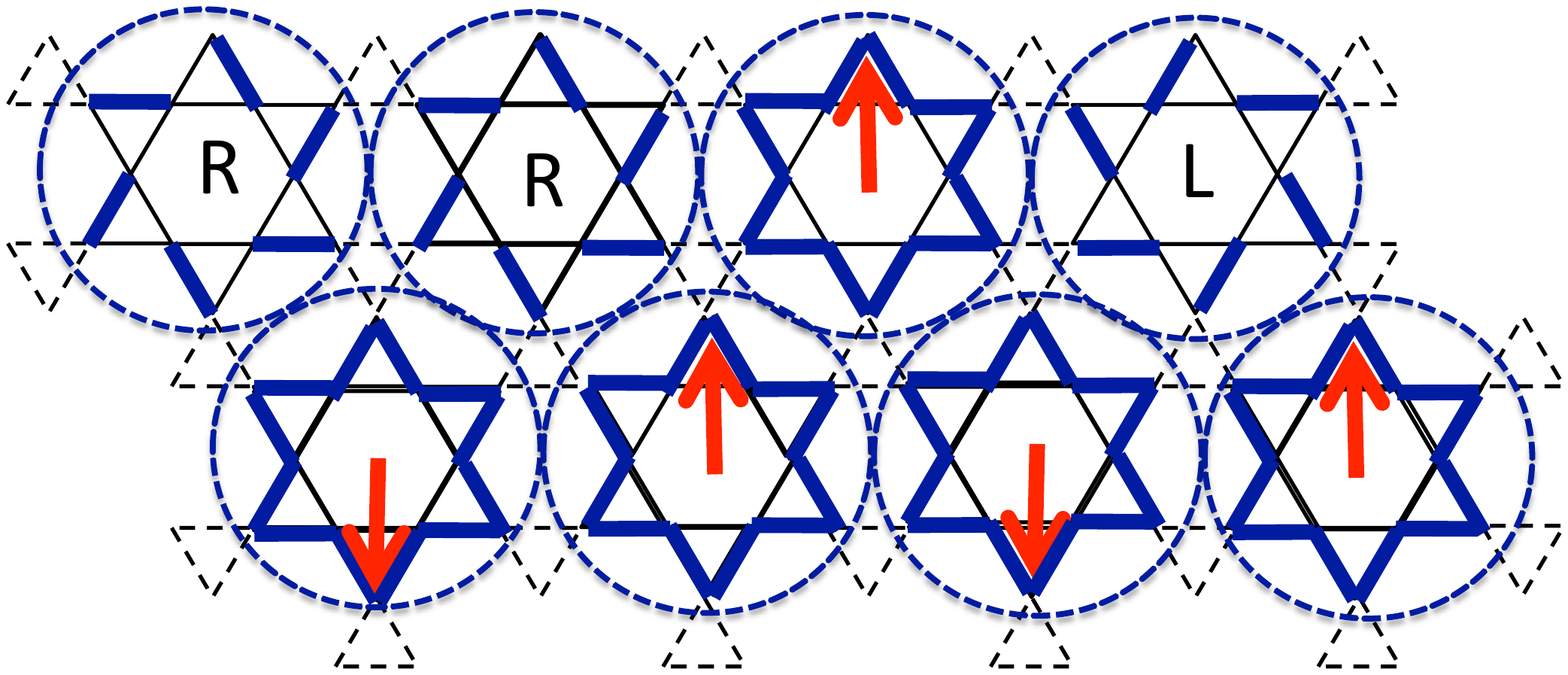}
\caption{\label{core-blocking}The CORE-blocking scheme on  the Kagom\'e lattice. R and L denote the two pinwheel ground states of the 12 site stars, and the 
arrows (pseudospins) denote the symmetrized Ising basis which spans the reduced Hilbert space of  $H^{CORE}$,  see Eq.~(\ref{pinwheels}).}
\end{figure}

This paper reports a surprising result: {\em The thermodynamic ground
  state appears to break reflection symmetries, and to possess
  two dimensional p6 chirality}, (-- not to be confused with ``spin chirality'' which also breaks time reversal symmetry~\cite{Messio}). Our  conclusion is obtained by  
Contractor Renormalization (CORE)~\cite{CORE-W} with 12-site stars
blocking, see Fig.~\ref{core-blocking}. 
the stars  scheme is found to reach sufficient accuracy
with range-3 interactions. This is evidenced by comparing ground state energies of the effective Hamiltonian $H^{CORE_3}$,
to high precision DMRG on large lattices. The small modulation  of bond energies is consistent, within our accuracy, with a translationally invariant singlet state as
deduced by DMRG~\cite{DMRG1,DMRG2}.

$H^{CORE_3}$  is  diagonalized on up to 27 stars
(effectively 324 Kagom\'e sites). The spectra exhibit doublet
degeneracies between states with opposite parity under reflection~\cite{epaps}. These signal a hitherto unexpected spontaneous  
symmetry breaking in the thermodynamic limit into a chiral ground state.  This chirality is
understood as the effect of three-star interactions.  Classical mean
field theory on the effective hamiltonian explains this symmetry breaking. A two-dimer chirality
order parameter  is defined on the microscopic Kagom\'e model.  We propose an experimental signature of this broken symmetry
in the phonon spectrum: a  splitting of symmetry-protected
degeneracy between two  zone center optical modes~\cite{Daniel}.

Finally, we add weak ferromagnetic next nearest neighbor interactions
$J_2$, and find that it  eliminates the chirality at $J_2
\approx  -0.1$.

{\em CORE procedure.} 
 Previous CORE calculations for the Kagom\'e model~\cite{CORE-A2,CORE-S} started with up-triangles blocking, and did not reach sufficient convergence at range 3.
Here we use much larger and more symmetric  blocks of 12 site (Magen David) stars which form a triangular superlattice.
In each star, we  retain just the  two  degenerate singlet ground states
 $|L_i\rangle$ and $|R_i\rangle$, depicted in Fig.~\ref{core-blocking} which form a pseudospin-1/2 basis:
 \bea
| \uparrow_i \rangle = \frac{1}{\sqrt{2+1/16}}(|R_i\rangle+ |L_i\rangle) ,\nonumber\\
| \downarrow_i \rangle = \frac{1}{\sqrt{2-1/16}}(|R_i\rangle- |L_i \rangle) .
\label{pinwheels}
\eea
Note that the two states are $C_6$-invariant, and have opposite parity under all $D_6$  reflections.

The  CORE effective Hamiltonian on a superlattice of size $N_s$ stars is defined by the cluster expansion,
\bea
 H^{CORE}  &=& \sum_{i=1}^{N_s} h^{(1)}_{i} +  \sum_{i_1 i_2} h^{(2)}_{i_1,i_2} +   \ldots +  \sum_{i_1,\ldots i_N} h^{(N_s)}_{i_1,\ldots i_{N_s}} ,\nonumber\\
h^{(n)}_{\alpha} &\equiv & H^{(n)}_\alpha   -  \sum_{m<n} \sum_{ \beta(m)\in \alpha(n)    }   h^{(m)}_{ \beta} .
\label{cluster}
\eea 

Here $\beta(m)$ is a connected subcluster of size $m$ in a
cluster $\alpha(n)$ of size $n$ stars, and $h^{(n)}$ is defined to be
an interaction of range-$n$.  The operators $H^{(n)}_\alpha$ are
constructed by ED of Eq. (\ref{H}) on a Kagom\'e cluster $\alpha$: 
\be
H^{(n)}_{\alpha} = \sum_{\nu}^{2^n} \epsilon^\alpha_\nu |
\tilde{\Psi}^\alpha_\nu \rangle \langle \tilde{\Psi}^\alpha_\nu | \ee Here
$ (\epsilon^\alpha_\nu,\Psi^\alpha_\nu)$ are the exact $2^n$ lowest
singlet energies and wavefunctions.  The states $|\tilde{ \Psi}_\nu
\rangle$ are an orthogonal basis constructed by sequential projections
of $|\Psi^\alpha_\nu\rangle, \nu=1,2,\ldots 2^n$ onto the pseudospin
states. After projection, the states are orthogonalized sequentially
by using the Gram-Schmidt procedure.

If interactions of all ranges $n\le N_s$ are included, then $H^{CORE}$ has the identical low energy singlet spectrum as
Eq.~(\ref{H}) on the equivalent Kagom\'e lattice.   However, ED cost to compute  $h^{(n)}$ grows exponentially with $n$.  
Thus, the success of CORE depends on the ability to truncate the cluster expansion at feasible  $n$ while maintaining sufficient accuracy in the truncated Hamiltonian.

The  error in the ground state energy $\delta E^{CORE_n}_0$ can be computed by comparison to high-precision  
DMRG on large lattices with $m>n$ stars. This error should be much smaller than the important  interactions in  $H^{CORE_n}$.

{\em Lattice translations.} Our choice of stars for the reduced Hilbert space {\em nominally} breaks lattice translational symmetry as seen in Fig.~\ref{core-blocking}.
The microscopic spin correlations are computed by  functional differentiation of the CORE ground state energy
with respect to source terms~\cite{CORE-A}. In principle one must compute the effective interactions to all ranges to restore full translational symmetry. 
Nevertheless, symmetry breaking artifacts  decrease with  the truncation range $n$.
We can therefore identify any spontaneous  translational symmetry breaking which significantly exceeds the  truncation error.

{\em CORE range 2.} 
We start with the lowest-order  truncation at range 2. The general form of the two-star interactions allowed by lattice reflection symmetries
is
\be
H^{CORE_2}  = Nc_0 +h \sum_i  \sigma^z_i  +  \sum_{\langle ij\rangle} J^\alpha\sigma_i^\alpha\sigma_j^\alpha  ,
\label{CORE2}
\ee 
where $i$ labels sites, $\langle ij \rangle$  nearest neighbor bonds on the triangular lattice. $\sigma^\alpha, \alpha=x,y,z$ are Pauli matrices.

The  parameters derived  from the lowest 4 eigenstates of 24 spins, are computed by Lanczos algorithm, and listed in Table \ref{Table1}.
It is instructive to compare the ED parameters to the second order perturbation theory (PT) in the inter-star bonds, as was calculated  by
Syromyatnikov and  Maleyev~cite{SM}. 
\begin{table} 
\begin{tabular}{|c|c|c|c|c|c|}
\hline
& $c_0$ & $h$   & $J^x$   & $J^y$ & $J^z$  \\ \hline
ED &  -6.26391  & 0.13818 &  0.00713 & -0.00105& -0.00045  \\ \hline
PT& -5.268  & 0.046  & 0  & -0.00025 & -0.00175 \\ 
\hline
\end{tabular} 
\caption{\label{Table1}Parameters of CORE range-2 Hamiltonian,  by Exact Diagonalization, and second order Perturbation Theory~\cite{SM}.}
\end{table}
Second order PT in the connecting bonds is not very accurate when connecting bonds have exchanges equal to  $1$.  
For example,  PT misses the important $J^x$ interactions. The dominant interaction  of $H^{CORE_2}$ is the field  $h=0.138$, which would yield in the thermodynamic system a ferromagnetic  ground state polarized in the $|\downarrow\rangle$ direction.  In terms of Kagom\'e  spins, 
the ground state would be a product of {\em antisymmetric} superposition of pinwheel states,
with local $\uparrow $  fluctuations generated by the $xx,yy$ terms.
 
Within CORE$_2$, the connecting bonds energy  is $E_{inter}= -0.21283$, versus the intra-star bonds at $E_{intra}= -0.2225$. Interestingly,
the modulation is already diminished from 100\%$\to$4.3\% with range-2 interactions.
         
{\em How accurate is $H^{CORE_2}$ ?  } 
Unfortunately, not enough. The exact ground state energy/site of ${\cal H}$ for 36 sites  is $E^{ED}_0$=-0.41276 while the CORE$_2$  energy/site for 
three stars is $E^{CORE_2}_0$=-0.4277.
The error of $-0.0149$ is quite large relative to the important terms in  $H^{CORE_2}$. Hence we cannot trust the truncation, and longer range interactions
are needed. 

{\em CORE range 3}.
To obtain $H^{CORE_3}$ we compute the  interactions $h^{(3)} $  on the three star triangular cluster \cite{comment1}. This required ED of Eq.~(\ref{H}) of 36 spins with open boundary conditions (OBC). For verification, we ran both a standard Lanczos routine on a supercomputer,  and the memory-economical Lanczos-SVD  routine \cite{LSVD}  on a desktop computer.  Adding  contributions from ranges 1-- 3  we obtain the following effective hamiltonian
\bea
H^{CORE_3}
&=&   N c_0+ \sum_i     h \sigma^z_i   + \sum_{\langle ij \rangle,\alpha }    J_\alpha \sigma^\alpha_i  \sigma^\alpha_j  \nonumber\\
&&
+   \sum_{\langle ijk\rangle_\Delta,\alpha }   J_{z\alpha\alpha}  \sigma^z_i   \sigma^\alpha_{j}  \sigma^\alpha_k   ,
\label{HC3}
\eea
where ${\langle ijk\rangle_\Delta}  $ label nearest neighbor triangles on the triangular lattice. We do not list terms that cancel in the 
superlattice summation with Periodic Boundary Conditions (PBC).

\begin{table}
\begin{tabular}{|c|c|c|c|c|}
\hline
$J_2$ & $J_2=$ 0  (Kagom\'e) & $J_2=$ +0.1  & $J_2=$ -0.1  \\ \hline 
$c_0$ & -5.24629 & -5.17068 & -5.48631 \\ \hline
$h$ & -0.069224  & 0.059323  & -0.362797 \\ \hline
$J_x$ & -0.009028 & -0.015421 & 0.001123 \\ \hline 
$J_y$ & -0.011879 & 0.001832  & -0.017699  \\ \hline
$J_z$ & 0.021056  & 0.003686  & 0.020141 \\ \hline
$J_{zxx}$ & -0.027920 & -0.019649 & -0.009524  \\ \hline 
$J_{zyy}$ & 0.004550  & -0.004749  & 0.002394  \\ \hline 
$J_{zzz}$ & 0.000660  & -0.001410  & 0.010495   \\
\hline
\end{tabular}\caption{Interaction parameters of CORE range 3, with three values of $J_2$.\label{Table2}}
\end{table}
The $H^{CORE_3}$  truncation error is estimated by comparing its ground state energy on clusters of sizes $N>3$  to high precision DMRG~\cite{White92,DMRGdetails}. From 
Table~\ref{Table3} we find that the truncation error is very satisfactory  $< 0.004 $ per site.  If we extrapolated CORE$_3$ ground state energy to the thermodynamic limit,
we get $ -0.447$ which underestimates  the extrapolated DMRG result  $-0.439$~\cite{DMRG2} by  at most $-0.008$ per site. 
We see in  Table~\ref{Table2} for comparison, that  the values of the dominant terms  $h$, $J_{zxx}$ and $J_z$ (multiplied by number of bonds/site), are significantly larger 
than this error.  {\em Thus, we believe  that CORE truncation at range 3 is sufficiently accurate to  obtain  the correct thermodynamic phase.}
\begin{table}
\begin{tabular}{|c|c|c|c|}
\hline
number of stars & $E_0^{CORE_3}$&   $E^{DMRG}_0$ & Error \\ \hline 
$2\times2 $ &  -0.418452 &  -0.417213 &  -0.001239  \\ \hline
$2\times3$ & -0.423953 &  -0.422336 &- 0.001617\\ \hline
$3\times4$ & -0.431150 & -0.428046 & -0.003104  \\\hline
$3\times5$ & -0.432688 & -0.429191 & -0.003497\\
\hline
\end{tabular}\caption{\label{Table3}Ground state energies per site  of $H^{CORE_3}$,  and comparison to DMRG on equivalent Kagom\'e clusters (with OBC).}
\end{table}

\begin{figure}
\includegraphics[width=\linewidth]{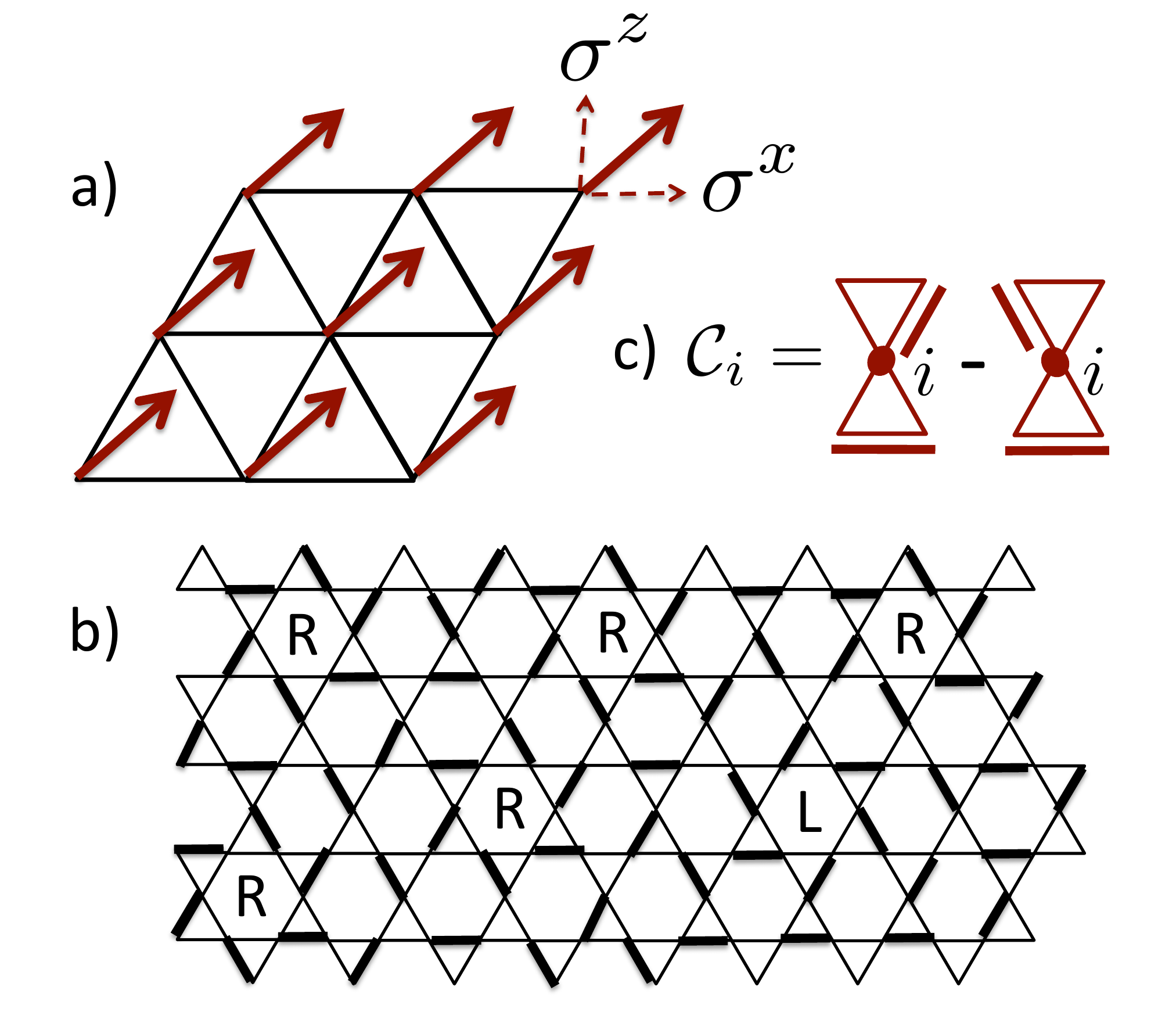}
\caption{a) The mean field ground state of $H^{CORE_3}$ exhibiting $\langle\sigma^x\rangle>0$ order, which corresponds to two dimensional chirality.
 b) A typical singlets configuration in the  corresponding ground state of the Kagom\'e lattice. Notice that there is no translational order, but that there are
 more  pinwheel configurations  of $|R\rangle$  than   $|L\rangle$. c) The two dimer chiral order parameter defined in Eq.\ref{OP}.}
\label{chiralRVB}
\end{figure}

{\em $p6$-Chirality}.  The  ED spectrum of  $H^{CORE_3}$ is evaluated  on lattices of up to $N_s=27$  stars (324 Kagom\'e sites), with
PBC.  The most striking feature on  lattices larger than $N_s=9$,
 is the emergence of ground state degeneracy of two singlets with opposite parity under reflections.
In the pseudospin representation,  even (odd) parity states include only an even (odd) number of stars with  antisymmetric $|\downarrow\rangle$ states.
These degeneracies signal a spontaneous reflection  symmetry breaking $p6m \to  p6$ in the thermodynamic limit.  

\begin{table}
\begin{tabular}{|c|c|c|}
\hline
$J_2$ & $M_z^{MF} $ &   $M_z^{ED}$  \\ \hline 
0.0  &  0.2647 & 0.2257      \\ \hline
+0.1& 0.1390&   0.1476 \\ \hline
-0.1& 0.5   &   0.4999  \\ \hline
\end{tabular}\caption{\label{Table4}Effect of next nearest neighbor couplings on the ground state $z$-polarization. }
\end{table}

A Mean Field (MF) energy of  $H^{CORE_3}$ in
spin-1/2 coherent states $|\bOm_i\rangle$ is
\bea
E^{MF}   =  N c_0 +h \sum_i \cos\theta_i   +     \sum_{\langle ij\rangle,\alpha} J_\alpha\bOm^\alpha_i\bOm^\alpha_j ,\nonumber\\
  + \sum_{\langle ijk\rangle,\alpha }J_{z\alpha\alpha}  \cos\theta_i \bOm^\alpha_j\bOm^\alpha_k,
\eea
where   $\bOm_i =(\sin\theta_i  \cos\phi_i ,\sin\theta_i \sin\phi_i ,\cos\theta_i )$.
In Table \ref{Table2} we see that for $J_2=0$,  the dominant couplings are the  field $h$, and  the 
$J_z$ and  $J_{zxx}$ exchanges.  The last coupling  is responsible for the chiral symmetry breaking, as it pulls  the spins in the $\pm \hat{x}$ direction.

Minimizing $E^{MF}$, we find a ferromagnetic state depicted in Fig.~\ref{chiralRVB}(a). 
The $z$-polarization $M^{MF}_z=\half \cos\bar{\theta}$ is compared to ED in Table \ref{Table4}.
For $J_2=0$,  we find that the chirality order is substantial with $\half \sin\bar{\theta} =0.424$ by MF, and $\half \langle\sigma^x\rangle=0.397$ by ED.

In Fig.~\ref{chiralRVB}(b) we  depict a typical dimer configuration which
contributes to the $p6$ - chiral RVB state.   One can see the predominance of $R$  pinwheel chirality over $L$. 
The most local order  parameter for this chirality is the two dimer correlation depicted in Fig.~\ref{chiralRVB}(c),
\be
\cC_{i} = \sum_{\bd} \left( \cS_\bd \cS_{\bfeta^r(\bd)} -  \cS_\bd \cS_{\bfeta^l(\bd)}\right) ,
\label{OP}
\ee
where the dimer singlet projectors are
\be
\cS_\bd = 1/4-\bS_{\bd_1} \cdot \bS_{\bd_2} ,
\ee
and  $\bfeta^r(\bd)$ ($\bfeta^l(\bd)$) is the bond emanating from $i$  at angle $\pi/3$ ($2\pi/3$) relative to the  dimer bond opposing $i$. 
The two terms in $\cC$ measure parts of pinwheels of opposite chirality.  

{\em Translational symmetry.} At range 3, the energy of internal triangles $E_{\Delta_{inter}}=-0.686$, and connecting triangles is $E_{\Delta_{intra}}= -0.665$  (depicted by solid and dashed lines respectively in Fig.~\ref{core-blocking}).     This 
relative modulation of about 3.0\% lies within the truncation error. Thus we can affirm that
CORE$_3$ ground state is consistent with \emph{translational  invariance}
in agreement with DMRG~\cite{DMRG1, DMRG2}.

{\em Singlet Excitations}. In the 27 star lattice, the lowest singlet excitation above the two degenerate ground states   is $\Delta E_{S=0} = 0.28$, which has
a non zero wavevector. This excitation gap does not vary much with lattice size. Within the pseudospin Hamiltonian, it can  be understood as  
a local spin flip from the ferromagnetic ground state. We note that the singlet gap is  slightly higher than two $S=1$ magnons at energies
$E_{S=1}=0.13$.  This conclusion differs from that obtained by ED on  36 site PBC, which found a large number of
singlets below the spin gap~\cite{EDlhuillier}. Since our effective Hamiltonian describes excitations on much larger lattices, we are inclined to associate these low singlets
with the  smaller PBC lattice geometry.

\begin{figure}
\includegraphics[width=\linewidth]{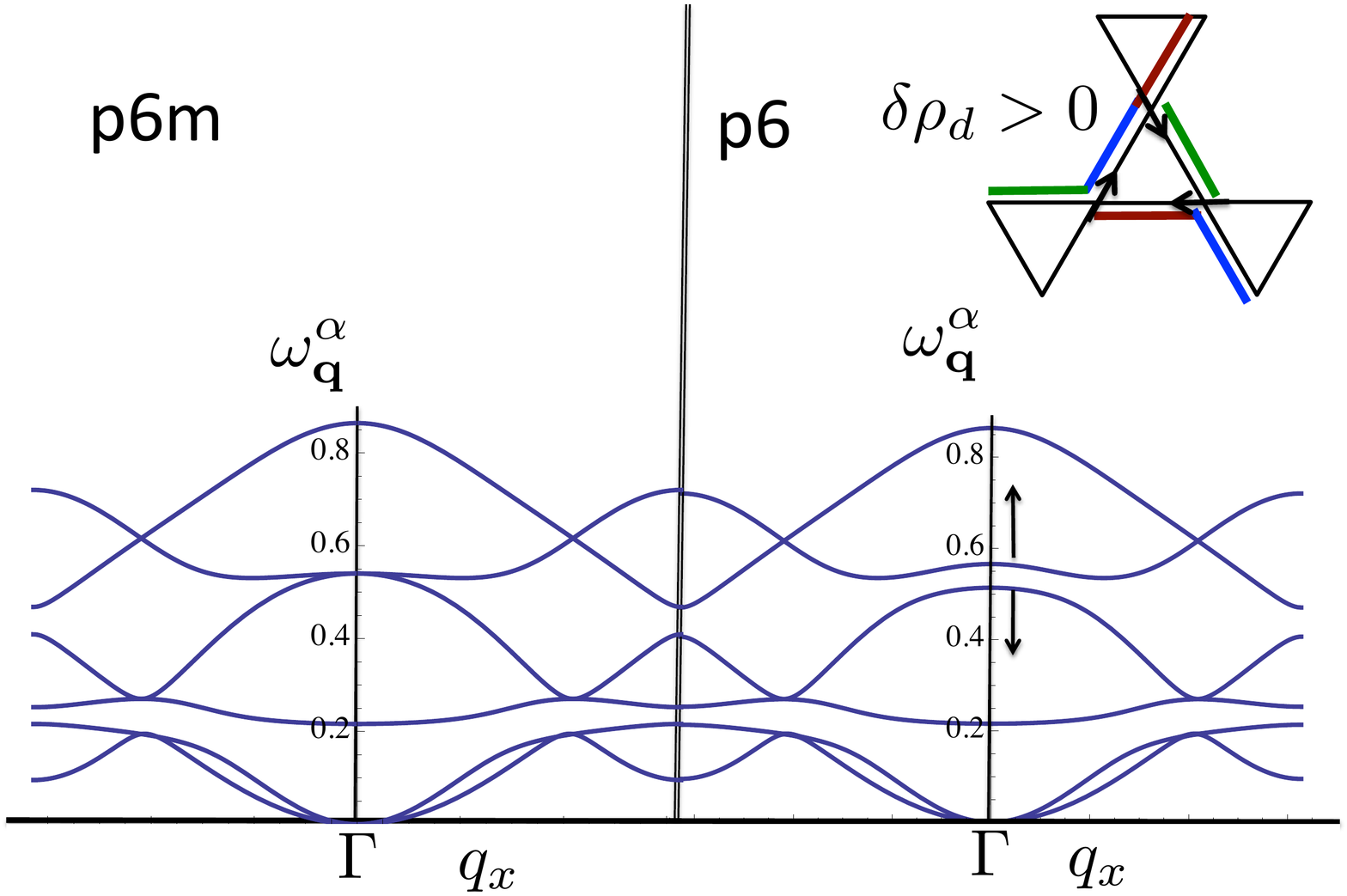}
\caption{ Kagom\'e phonon spectra in $p6m$ phase, and $p6$ - chiral phase, calculated within  a nearest neighbor spring constant model given in Ref.\cite{phonons}.
In the top right, the dimer chiral correlations induce a linear coupling between
excess dimer density $\delta\rho_d$ and chiral ionic displacements, as depicted by the arrows. This adds a chiral term to the dynamical matrix which splits the degeneracy
of the optical phonons at the zone center. }
\label{fig:phonons}
\end{figure}

{\em Experimentally,}  fluctuating two-dimer correlations are tricky to observe directly. Fortunately,  real compounds have sizable magneto-elastic coupling
between the ions and the dimer singlets. 
While, on average, dimer density and  bond lengths are uniform in the RVB state, 
{\em dimer density fluctuations}, $\delta\rho_d$ governed by the characteristic singlet energy scale,  are linearly coupled to the ionic displacements.     
In Fig.~\ref{fig:phonons}, the effect of  a temporary excess of dimers on a triangle is shown.
In the chiral phase, imbalance between the left and right bonds emanating out of the triangle produces a chiral force on the ions as depicted  by the
arrows. Integrating out the dimer density fluctuations 
results in a chiral perturbation to the phonon dynamical matrix~\cite{epaps}.  By symmetry~\cite{Daniel},  the degeneracy 
between two  optical modes is removed at the zone center, as shown in Fig.~\ref{fig:phonons}.

{\em Finite $J_2$}. We have  added  next nearest neighbor interactions with coupling $J_2$ to Eq. (\ref{H}), and calculated the parameters of $H^{CORE_3}$, as shown in Table \ref{Table2}. For $J_2= 0.1$, we find 
the same doublet degeneracies, and chirality,  as for the pure model $J_2=0$~\cite{BalentsZ2}. In contrast, for a weak  negative $J_2=-0.1$  the spectrum changes dramatically: The doublets are removed, and the ground state is  fully polarized in the $\uparrow$ direction.
The precise nature of this phase  has not been yet explored. Interestingly, we
notice that  in  proximity to the parameters of Table~\ref{Table2},  one finds the Ising antiferromagnet in field. Its ground state contains
ferromagnetic hexagons with reversed spins in their center. It represents the  Hexagonal Valence Bond  Solid state, previously shown to have low variational energies~\cite{HVBC},  and proposed for $J_2 \simeq -0.1$~\cite{J2}.

{\em Summary.} Using CORE we arrived at an effective Hamiltonian, whose accuracy was determined to be sufficiently high so as to trust its predictions for the thermodynamic
limit.  Its ground state is consistent with a translationally invariant RVB phase, but with broken p6 chiral symmetry. A 2-dimer chiral order parameter is defined, which 
may be numerically explored on large latices.  Experimentally,  it may  be
detected by splitting of optical phonon degeneracy.

{\em Acknowledgements.} We thank Andreas L\"auchli, Daniel Podolsky and Didier Poilblanc,  for useful discussions. AA and SC acknowledge the hospitality of the Aspen Center for Physics and KITP at Santa Barbara,   supported by the NSF Grants  PHY-1066293  and  PHY-1125915. Funding  from U. S. DOE, Contract No.~DE-AC02-76SF00515, Israel Science Foundation
and U.S. -- Israel Binational Science Foundation are acknowledged. Numerical simulations were performed at CALMIP.

 

\begin{thebibliography}{99}
\bibitem{Daniel} We are indebted to Daniel Podolsky for suggesting to us this idea.
\bibitem{DMRGdetails} We use open boundary conditions on these clusters, and keep up to to get a discarded weight smaller than $10^{-5}$. Note that these variational DMRG energies give exact upper bounds on the ground-state energy~\cite{DMRG1}.
\end{thebibliography}
\end{document}


\title{Supplement: p6-Chiral Resonating Valence Bonds in the Kagom\'e Antiferomagnet}
\author{Sylvain Capponi$^{1}$}
\author{V. Ravi  Chandra$^{2,3}$}
\author{Assa Auerbach$^3$}
\author{Marvin Weinstein$^4$}
\affiliation{$^1$ Laboratoire de Physique Th\'eorique, Universit\'e de Toulouse and CNRS, UPS (IRSAMC), F-31062, Toulouse, France}
\affiliation{$^2$ Physics Department, Technion, Haifa 32000, Israel}
\affiliation{$^3$ National Institute of Science Education and Research (NISER),
IOP Campus, Bhubaneswar, 751005, India}
\affiliation{$^4$ SLAC National Accelerator Laboratory, Stanford,  CA 94025, USA.}
\date{\today}
\begin{abstract}
Here we provide details of the exact diagonalization of $H^{CORE_3}$ on 27 stars, and the effects of the chiral order parameter on the optical phonons
of the Kagom\'e ions.\end{abstract} 
\maketitle

\section{Exact Diagonalization of 27 stars}
\begin{figure}[!htb]
\includegraphics[width=\linewidth,clip]{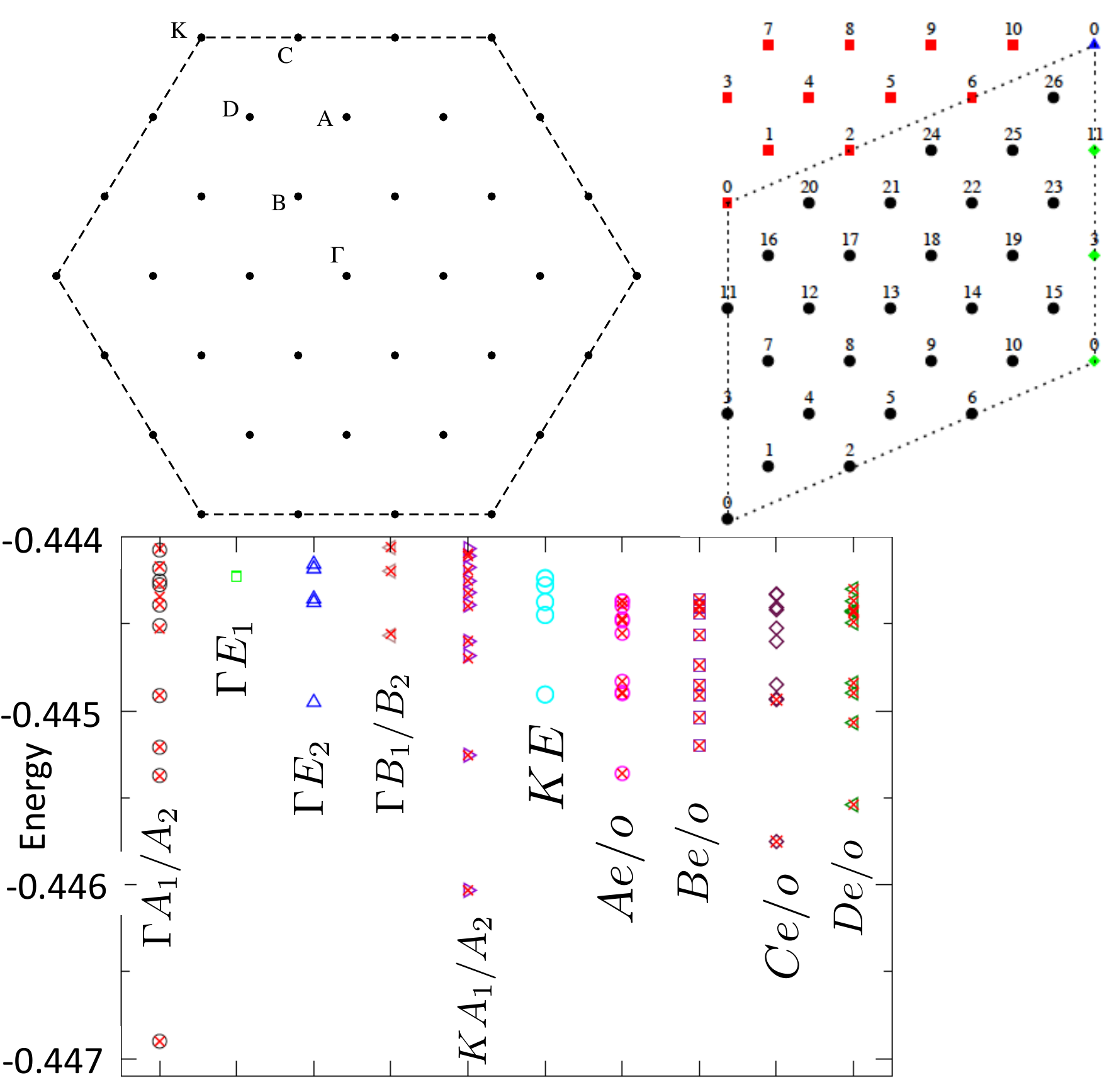}
\caption{Top: a cluster of 27 sites on the triangular superlattice, and its wavevectors in the Brillouin Zone Spectrum.
Bottom: The singlet spectrum of $H^{CORE_3}$ on this cluster (equivalent to 324 Kagom\'e spins). 
Notice the nearly exact doublet degeneracies, 
including the ground state doublet at $\Gamma A_1, \Gamma A_2$. 
These degeneracies  signal $p6m \to p6$ spontaneous  symmetry breaking in the thermodynamic limit.\label{fig:27MD}}
\end{figure}

In Fig.~\ref{fig:27MD} we plot the exact spectrum of 27 stars with periodic boundary conditions. Since this cluster has all translations and the full C$_{6v}$ point group symmetry, we can label eigenstates according to their symmetry quantum numbers. The 6 inequivalent momenta are shown in the plot, and for each of them, we can make use of the little group, i.e. the reduced symmetry group that leaves the momentum invariant: at $\Gamma$ point, the little group is C$_{6v}$, at $K$ point C$_{3v}$, while A, B, C, and D momenta only have a single $Z_2$ reflection. Therefore, we can label few low-energy states according to the irreducible representations of these groups. 

\section{Coupling of the Chiral Order parameter and the Phonon spectrum}
{\em p6m Model}

A simple model for the phonon spectrum of the planar Kagom\'e was worked out by L. Zhang {\em et al.} in Ref.~\cite{phonons}.
The two dimensional ionic dispacements $\bu_i$ are labelled by $i=$1= bottom left, 2= bottom right, and 3= top site. The nearest neighbor force constant matrix in the $xy$ basis is
\be
K_x = \left( \begin{array}{cc} K_L & 0\\ 0& K_T\\
\end{array} \right)
\ee
where we $K_L$ and $K_T$ are longitudinal and transverse force constants respectively.  We follow Ref.~\cite{phonons} and take $K_T=K_L/4$.
A  rotation is defined:
\be
U= \left( \begin{array}{cc} \cos(\pi/3)  & -\sin(\pi/3)\\ \sin(\pi/3)& \cos(\pi/3)\\
\end{array} \right)
\ee
with which we express the rotated force constant matrices 
\be
K_{12} =K_x, ~~~K_{13}= U K_x U^\dagger,~~~K_{23}= U^\dagger K_x U
\ee
These matrices define blocks in the $6\times 6$ spring constant matrices
\bea
K_0&=&  \left( \begin{array}{ccc} 
2(K_{12} + K_{13})& -K_{12}& -K_{13}\\ 
 -K_{12}& 2(K_{12} + K_{23})& -K_{23}\\ 
-K_{13}& -K_{23}& 2(K_{13} + K_{23})
\end{array} \right)\nonumber\\
K_1&=&  \left( \begin{array}{ccc} 
0&0&0\\ 
 -K_{12}&0&0\\ 
0&0&0
\end{array} \right)~~~~~
K_2 =   \left( \begin{array}{ccc} 
0&0&0\\ 
 0&0&0\\ 
-K_{13}&0&0
\end{array} \right)\nonumber\\
K_3&=&  \left( \begin{array}{ccc} 
0&0&0\\ 
 0&0&0\\ 
0&-K_{23}&0
\end{array} \right)~~~~~
K_4=  \left( \begin{array}{ccc} 
0&-K_{12}&0\\ 
 0&0&0\\ 
0&0&0
\end{array} \right)\nonumber\\
K_5&=&  \left( \begin{array}{ccc} 
0&0&-K_{13}\\ 
 0&0&0\\ 
0&0&0
\end{array} \right)~~~~~
K_6=  \left( \begin{array}{ccc} 
0&0&0\\ 
 0&0&-K_{23}\\ 
0&0&0
\end{array} \right)
\label{Kmatrix}
\eea
The dynamical matrix in $\bq$-space is 
\bea
K(\bq) &=& K_0 + K_1 e^{i q_x} + K_2 e^{i( {q_x\over 2} + {\sqrt{3}q_y\over 2} )} \nonumber\\
&&+  K_3 e^{i( -{q_x\over 2} + {\sqrt{3}q_y\over 2} )} + K_4 e^{-i q_x} \nonumber\\
&&+  K_5 e^{i( -{q_x\over 2} - {\sqrt{3}q_y\over 2} )} +K_6 e^{i( {q_x\over 2} - {\sqrt{3}q_y\over 2} )}   \nonumber\\
\label{Kq}
\eea

\begin{figure}[!htb]
\includegraphics[width=\linewidth]{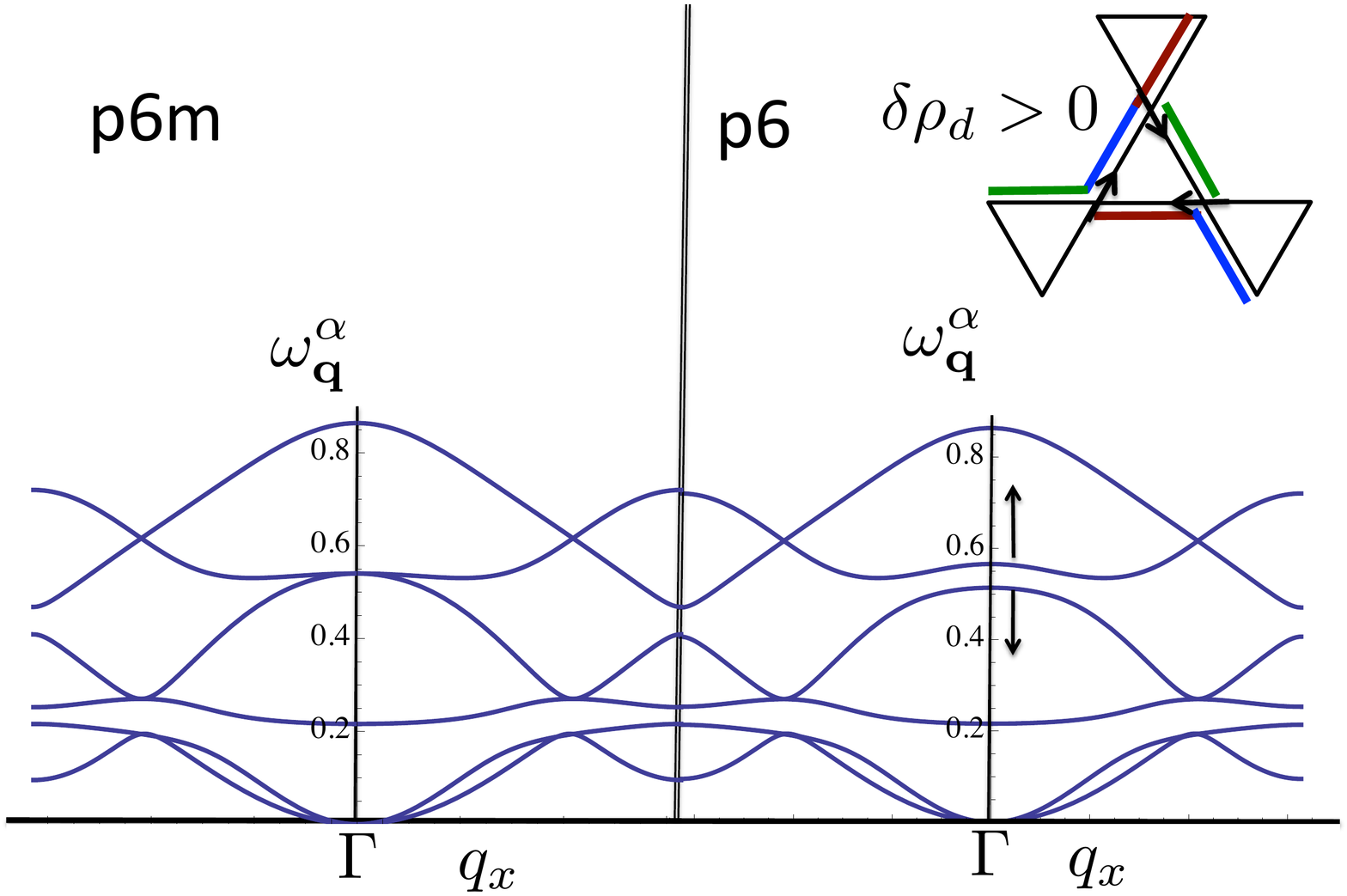}
\caption{Kagom\'e phonon spectra in p6m and p6-Chiral phases, calculated within  a nearest neighbor spring constant model given in Ref.~\cite{phonons}.
In the top right, the dimer chiral correlations induce a linear coupling between
excess dimer density and chiral ionic displacements, as depicted by the arrows. This adds a chiral term to the dynamical matrix and splits the degeneracy
of the optical phonons at the zone center.\label{fig:phonons} }
\end{figure}

{\em Chiral Symmetry Breaking RVB --}
A magneto-elastic coupling ${\delta J\over \delta u}$ 
exerts a force on the   atoms on either side of a singlet dimer toward each other.
However, the RVB phase has a uniform average density of dimers on all bonds. There are however dynamical fluctuations  
of  the dimer density which can be modelled by  a harmonic Lagrangian
\be
\CL_{d} = \half \int d^2x \half  (\partial_t^2 -  \omega_d^2) \delta\rho_d^2(\bx)
\ee
where $\omega_d$ is of the order of the singlet excitation energies.

Let us consider a RVB phase with positive chirality given by $\langle \cC\rangle >0$. An instantaneous excess  dimer density  on the up-triangle 
is  shown in Fig.~\ref{fig:phonons}. (In any dimer covering, some fraction of the triangles have no dimers, and thus lower than the average density).
 Due to right chirality, the neighboring triangles' dimers directions are biased
to  emanate rightward relative to the opposite edge.   As a consequence, each corner of our triangle feels a net "pull" along the direction of its 
(on the average) uncompensated bond. This amounts to an effective coupling
\be
\CL_{d-ion} = -\gamma |\langle \cC\rangle|  \int d^2x  \delta\rho_d^2(\bx) \bv^r \cdot \bu(\bx)
\ee
where $\bu$ is the ion displacements vector, and $\bv^r$ is given as the 6-dimensional clockwise displacements,
\be 
\bv^r = {1\over \sqrt{3}} (\cos({\pi\over 3} ), \sin({\pi\over 3}), -1, 0, \cos({\pi\over 3}) , -\sin({\pi\over 3}) \nonumber\\
\ee 
By integrating out the gapped dimer density fluctuations, a chiral correction
to the quadratic  dynamical matrix is obtained
\be
D^{chiral} (\bq)= D(\bq) - {\gamma^2  |\langle \cC\rangle|^2   \over \omega_d }  \bv^r \bv^r 
\ee
This correction lifts the degeneracy of the optical phonons at $\bq=0$ as shown in Fig.~\ref{fig:phonons}.

